\documentclass[journal]{IEEEtran}[12pt]
\usepackage[table]{xcolor}
\usepackage{psfrag}
\usepackage{acronym}
\usepackage{etoolbox}
\usepackage{wrapfig}
\usepackage{epsfig}
\usepackage{colortbl,color}
\usepackage{amsmath}
\usepackage{amssymb}
\usepackage{mathabx}
\usepackage{enumerate}
\usepackage{bbm}
\usepackage{algorithm}
\usepackage{algorithmic}
\usepackage{lipsum,amsmath,multicol}
\usepackage{stfloats}
\usepackage{arydshln} 
\usepackage{amsfonts}
\usepackage{dsfont}
\usepackage{upgreek}
\usepackage{pgfplots} 
\usepackage{soul}
\usepackage{tabularx}
\usepackage{booktabs}
\usepackage{multirow}
\usepackage{graphicx}
\usepackage{pgfplots}
\usepackage{url}

\usepackage{cite}
\usepackage{verbatim}
\acrodef{1G}{first generation}
\acrodef{2G}{second generation}
\acrodef{3G}{third generation}
\acrodef{3GPP}{3rd Generation Partnership Project}
\acrodef{4G}{fourth generation}
\acrodef{5G}{fifth generation}
\acrodef{6G}{sixth generation}
\acrodef{AI}{artificial intelligence}
\acrodef{AoA}{angle of arrival}
\acrodef{BER}{bit error rate}
\acrodef{CPL}{central perpendicular line}
\acrodef{BS}[BS]{base station}
\acrodef{CRLB}{Cramer-Rao lower bound}
\acrodef{MME}{model-mismatch error}
\acrodef{MCRLB}{Mismatched Cramer-Rao lower bound}
\acrodef{CS}{compressive sensing}
\acrodef{CID}{cellular ID}
\acrodef{DCS}{digitally controllable scatterer}
\acrodef{DMA}{dynamic metasurface antenna}
\acrodef{EM}{electromagnetic}
\acrodef{FIM}{Fisher information matrix}
\acrodef{KKT}{Karush–Kuhn–Tucker}
\acrodef{KPI}{key performance indicator}
\acrodef{LoS}{line-of-sight}
\acrodef{CSI}{channel state information}
\acrodef{DoF}{degrees of freedom}
\acrodef{EMF}{electromagnetic fields}
\acrodef{E-CID}{enhanced-CID}
\acrodef{ES}{emergency services}
\acrodef{IoT}{internet-of-things}
\acrodef{i.i.d.}{independent and identically distributed}
\acrodef{LIS}{large intelligent surface}
\acrodef{LPP}{LTE Positioning Protocol}
\acrodef{LMMSE}{linear minimum mean square error}
\acrodef{ZF}{zero-forcing}
\acrodef{mm-waves}[MM-Waves]{millimeter-waves}
\acrodef{MLE}{maximum likelihood estimator}
\acrodef{SISO}{single-input single-output} 
\acrodef{MAP}{maximum a posteriori} 
\acrodef{MIMO}{multiple-input multiple-output}
\acrodef{MS}[UE]{user equipment}
\acrodef{Multi-RTT}{multi round trip time}
\acrodef{NLoS}{non-line-of-sight}
\acrodef{NR}{new radio}
\acrodef{NRPPa}{NR Positioning Protocol A}
\acrodef{OFDM}{orthogonal frequency-division multiplexing}
\acrodef{RFID}{radio-frequency identification}
\acrodef{RIS}{reconfigurable intelligent surface}
\acrodef{RSS}{received signal strength}
\acrodef{SDS}{software defined surface}
\acrodef{SER}{symbol error rate}
\acrodef{SINR}{signal-to-interference-plus-noise ratio}
\acrodef{SRE}{smart radio environment}
\acrodef{SNR}{signal-to-noise ratio}
\acrodef{SNS}{spatial non-stationarity}
\acrodef{SWM}{spherical wavefront model}
\acrodef{SVD}{singular value decomposition}
\acrodef{SL}{service level}
\acrodef{POA}{phase-of-arrival}
\acrodef{ToA}{time-of-arrival}
\acrodef{TDoA}{time difference-of-arrival}
\acrodef{COA}{curvature-of-arrival}
\acrodef{AoD}{angle-of-departure}
\acrodef{RSSI}{received signal strength indicator}
\acrodef{SLAM}{simultaneous localization and mapping}
\acrodef{CRLB}{Cramér-Rao lower bound}
\acrodef{GDOP}{geometric dilution of precision}
\acrodef{PEB}{position error bound}
\acrodef{pdf}{probability density function}
\acrodef{PE}{positioning element}
\acrodef{PDoA}{phase difference-of-arrival}
\acrodef{IPS}{smartphone-centered indoor positioning systems}
\acrodef{PRS}{position reference signals}
\acrodef{OEB}{orientation error bound}

\acrodef{RARE}{Rank-Reduced }
\acrodef{RMSE}{root mean square error}
\acrodef{UAV}{unmanned aerial vehicle}
\acrodef{ULA}{uniform linear array}
\acrodef{UTDoA}{uplink time difference-of-arrival}
\acrodef{UAoA}{uplink angle-of-arrival}
\acrodef{eMBB}{enhanced mobile broadband}
\acrodef{RA}{resource allocation}
\acrodef{RF}{radio frequency}
\acrodef{QoS}{quality of service}
\acrodef{ESPRIT}{estimation of signal parameters via rotational invariance technique}
\acrodef{MUSIC}{multiple signal classification}
\acrodef{LS}{least squares}
\acrodef{LCS}{location service}
\acrodef{E-OTDOA}{enhanced observed time difference-of-arrival}
\acrodef{ELAA}{extremely large aperture arrays}
\acrodef{OTDOA}{observed time difference-of-arrival}
\acrodef{GPS}{global positioning system}
\acrodef{GSM}{global system for mobile communications}
\acrodef{GOSPA}{generalized optimal subpattern assignment}
\acrodef{PRS}{position reference signals}
\acrodef{UE}{user equipment}
\acrodef{UWB}{ultra-wide bandwidth}
\acrodef{XR}{Extended Reality}
\acrodef{VLC}{visible light communication}
\acrodef{VLP}{visible light positioning}
\acrodef{LiDAR}[Li-DAR]{light detection and ranging}
\acrodef{UHF}{ultra high frequency}
\acrodef{ZZB}{Ziv-Zakai bound}
\acrodef{MSE}{mean square error}
\acrodef{MLR}{mismatched likelihood ratio}

\acrodef{FFM}{far field model}
\acrodef{KL}{Kullback–Leibler divergence}
\acrodef{FF}{far-field}
\acrodef{NF}{near-field}

\title{Near and Far Field Model Mismatch: Implications on {6G} Communications, Localization, and Sensing}
\author{Ahmed~Elzanaty,
	~\IEEEmembership{Senior Member,~IEEE}, Jiuyu Liu,
	~\IEEEmembership{Student Member,~IEEE,} 
	Anna Guerra,
	~\IEEEmembership{Member,~IEEE,} \\
	Francesco Guidi,
	~\IEEEmembership{Member,~IEEE}, Yi Ma,
	~\IEEEmembership{Senior Member,~IEEE}, Rahim Tafazolli,
	~\IEEEmembership{Senior Member,~IEEE} 
	\thanks{A. Elzanaty (a.elzanaty@surrey.ac.uk), J. Liu, Y. Ma and R. Tafazolli are with the 5GIC \& 6GIC, ICS, University of Surrey, Guildford, UK. 
		A. Guerra and F. Guidi are with CNR-IEIIT, Italy}
	\thanks{This work is supported by the UK Department for Science, Innovation and Technology under the Future Open Networks Research Challenge project TUDOR (Towards Ubiquitous 3D Open Resilient Network). The views expressed are those of the authors and do not necessarily represent the project.
	}
}
\definecolor{aliceblue}{rgb}{0.94, 0.97, 1.0}
\definecolor{greenyellow}{rgb}{0.7, 0.9, 0.4}
\pgfplotsset{compat=1.16}

\begin{document}
	
	\maketitle
	
	\begin{abstract}
		The upcoming 6G technology is expected to operate in \ac{NF} radiating conditions thanks to high-frequency and electrically large antenna arrays. Although several studies have already addressed this possibility, it is worth noting that \ac{NF} models introduce higher complexity, the justification for which is not always evident in terms of performance improvements. 
		This article investigates the implications of the mismatch between \ac{NF} and \ac{FF} models concerning communication, localization, and sensing systems. Such disparity can lead to a degradation of performance metrics such as sensing and localization accuracy and communication efficiency. By exploring the effects of mismatches between \ac{NF} and \ac{FF} models, this study seeks to revolve around the challenges faced by system designers, offering insights about the balance between model accuracy and achievable performance. Finally, we conduct a numerical performance analysis to verify the impact of the mismatch between \ac{NF} and \ac{FF} models.
	\end{abstract}
	
	
	\acresetall
	\bstctlcite{IEEEexample:BSTcontrol}

	\section{Introduction}
	
	The next \ac{6G} of wireless communication is poised to become a convergence point for communication, localization, and sensing capabilities.  We refer to localization as the process of finding the 2D or 3D coordinates of a connected \ac{UE} by analyzing measurements from \acp{BS}. On the other hand, sensing resembles radar-like operations, including detecting and tracking passive objects. 6G will combine these three essential components, unlike previous wireless generations, creating a holistic and interconnected system. By seamlessly integrating these functionalities, 6G wireless technology aims to revolutionize various domains, including \ac{IoT} and autonomous systems. This convergence will enable unprecedented connectivity, precision, and situational awareness levels, empowering applications that require real-time data exchange, precise positioning, and intelligent sensing.
	
	One of the key features of \ac{6G} technology that distinguishes it from its predecessors is the use of high-frequency massive antenna systems. Consequently, most of the \acp{UE} will be in the \ac{NF} region of these electrically large antennas, introducing unique challenges in modeling and computational complexity associated with localization, communication, and sensing algorithms. 
	
	Indeed, accurate \ac{NF} models become essential to capture the intricacies of the communication and sensing environments and mitigate performance losses. However, using such models often comes with increased computational complexity, posing a trade-off between achieving precise localization, reliable communication, and accurate sensing while efficiently managing computational resources. 
	Notably, selecting an optimal model is particularly important for two main reasons: \textit{(i)} the model's accuracy can significantly impact the system performance; \textit{(ii)} the choice of model can affect the complexity and cost of the system.
	However, selecting the optimal model is not easy as multiple factors must be considered before deciding. First, no exact quantifiable boundaries exist between \ac{NF} and \ac{FF} because they depend on the geometry and system parameters. Moreover, this selection necessitates the assessment of various aspects, including the application, desired level of accuracy, system complexity, and cost. The Fraunhofer distance, commonly used as the boundary between \ac{NF} and \ac{FF}, fails to consider crucial factors such as the beam squint effect for wideband signals, transmission power, and \ac{AoA}. The reason is that its definition relies on the phase variation of a monochromatic wave across the length of the array \cite{HuiChenElzanaty:22,MinDai:23}.
	
	This paper investigates the disparity between \ac{NF} and \ac{FF} models and algorithms from a novel perspective of model mismatch in three pivotal service categories: communication, localization, and sensing. The proposed viewpoint illustrates the main differences between commonly adopted models and algorithms in array signal processing for \ac{NF} and \ac{FF} conditions. Then, we illustrate several metrics to quantify the mismatch and its implication on system performance. To the authors' knowledge, this is the first work considering localization, communication, and sensing from the \ac{NF} and \ac{FF} model mismatch perspective. More precisely, we present several methods to quantify the model mismatch and its impact on the algorithm design and system performance.
	
	\section{Near-Field Models and Algorithms for Localization and Communications}
	This section defines \ac{NF} models and discusses the necessity of having tailored \ac{NF} algorithms.
	
	\subsection{Definition and Characteristics} \label{sec:nearfieldfeatures}
	\ac{NF} models are mathematical descriptions that explain how electromagnetic fields behave when an antenna receiver, is situated close to a transmitter. The \ac{NF} region is divided into reactive and radiative (also known as the Fresnel region) parts. The reactive area is very close to the antenna and has complex, unpredictable relationships between electric and magnetic fields. \footnote{Even if the reactive near-field region is typically close to the antenna, some applications, like the \ac{RFID} one, have exploited such area for short-distance communications and power transfer.} 
	On the other hand, the radiative \ac{NF} region, which is affected by the operating frequency and the antenna aperture, is farther from the antenna compared to reactive \ac{NF} but still within the Fraunhofer distance \cite{bjornson2020power}.
	
	In this region, differently from the reactive zone, the EM waves experience a radiative propagation, like in the \ac{FF}, with a spherical wavefront and energy distribution influenced by the distance from the source.
	These models differ from \ac{FF} models, focusing on radiation properties at distances greater than the Fraunhofer distance. In the \ac{FF}, power decreases with the square of the distance. The wavefront can be approximated as planar, enabling simplified mathematical formulations and beamforming techniques. Nevertheless, this regime does not allow the power to be focused on precise points of the space.
	
	\ac{NF} models offer valuable advantages due to the availability of additional \ac{DoF}, finding use in various applications. They provide high-resolution data about electromagnetic fields, including detailed information on field distribution, polarization, and phase characteristics about the source, scatterer, and interfering \ac{UE} locations. This extra information aids precise design and optimization of communication, localization, and sensing systems. 
	
	Nevertheless, \ac{NF} models have some limitations as well. For instance, designing algorithms for the \ac{NF} while being in the \ac{FF} can pose ill-conditioned challenges, especially when utilizing exclusive \ac{NF} \ac{DoF}. As an example, in \cite{elzanaty2021reconfigurable} it is shown how transitioning from \ac{NF} to \ac{FF} results in a loss of information, thereby significantly reducing the accuracy of orientation estimation. Moreover, \ac{NF} models can be computationally demanding, particularly for complex geometries and sources, requiring sophisticated numerical methods and increased resources. This results in higher latencies, higher costs, and fewer insights than the simpler \ac{FF} models.
	
\begin{figure}
		\centering
		\includegraphics[width=0.90\linewidth,clip]{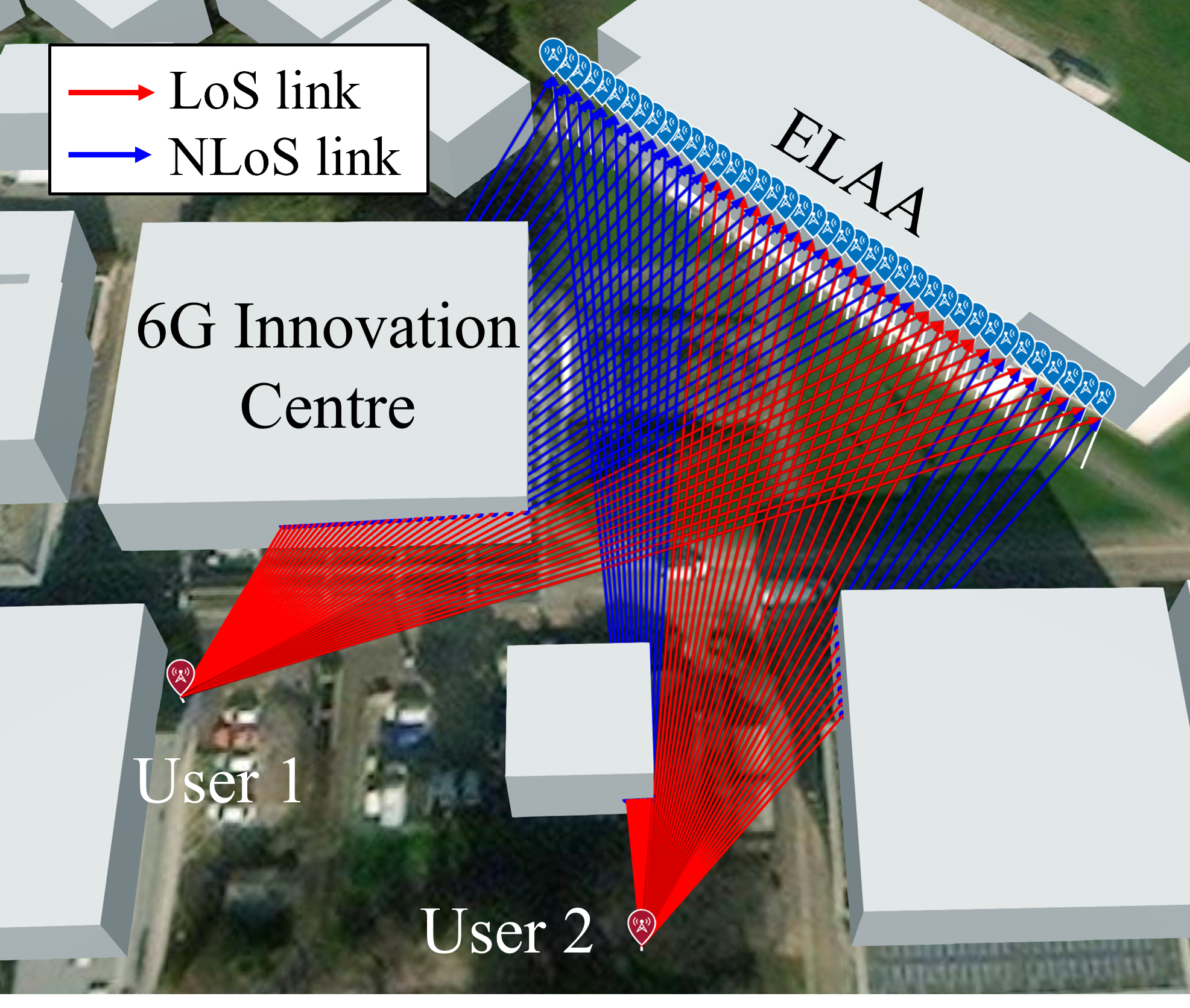}
		\caption{A practical example of an \ac{ELAA} channel with spherical wavefront and a mix of LoS and NLoS links. The spatial inconsistency on the network side introduces significant diversity in channel spatial distribution, characterized as spatial non-stationarity.} 
		\label{fig:enter-label2}
	\end{figure}

	Due to such considerations, it is crucial to select the proper field condition that balances complexity with \ac{QoS}. 
	As an example, Fig.~\ref{fig:enter-label2} illustrates that when \ac{ELAA} technology is implemented in \ac{MIMO} systems, user devices could be situated in \ac{NF}. 
	This leads to  \ac{SNS} in the wireless channel, a consequence of inconsistent path loss, shadowing, and LoS conditions that \ac{FF} models cannot capture. Therefore, the methods developed for communication, sensing, and localization depending on \ac{FF} models could exhibit sub-optimal performance in  \ac{NF} conditions.
	Conversely, while NF channel models might provide higher accuracy than FF models, this advantage becomes negligible in an FF environment where FF models already offer sufficient accuracy. 
	Moreover, opting for NF models in such scenarios could increase the computational burden due to their higher complexity. 
	Indeed, while the channel condition remains non-negotiable, the choice of the adopted model within the algorithm is subject to our understanding, which may not always align perfectly with the ground truth model. Therefore, if we have a prior distribution for the position, we can rely on Bayesian-based approaches or model regime-switching approaches to select the appropriate model \cite{9287507}.
	\subsection{Models for Array Signal Processing}
	To comprehend models in literature for applications like communication and sensing, let us consider a source-generated impinging wave on an antenna array. Signal representation hinges on the chosen model for the array manifold. Four models are presented, progressing from least to most intricate: \textit{(i)} \ac{FF}, \textit{(ii)} \ac{SWM}, \textit{(iii)} \ac{SWM} and \ac{SNS}, and \textit{(iv)} \ac{EMF}-based array manifolds.
	
	For the \ac{FF} assumption, vector phase components rely solely on angular information, \ac{AoA}, excluding the influence of the range. Phase shifts between antennas depend solely on spacing and \ac{AoA}, assuming a planar wavefront \cite[Eq. 9]{friedlander2019localization}.
	
	The \ac{SWM} model, a \ac{NF} variant, introduces phase shifts determined by array geometry, \ac{AoA}, and range from source to antennas. Signal strength remains uniform across antennas, yet the spherical wavefront approximates actual wave behavior in the Fresnel region \cite[Eq. 13]{friedlander2019localization}. \ac{SWM} with \ac{SNS} is similar to the \ac{SWM} model, but the signal strength at each antenna varies based on distance from source to that antenna \cite[Eq. 6]{HuiChenElzanaty:22}.\footnote{Note that \ac{SNS} stands as an independent feature regardless of whether a planner or spherical wavefront is experienced.}
	
	While these models are widely used in signal processing for arrays, they disregard source radiation patterns and antenna coupling. \ac{EMF}-based manifolds address these gaps, accurately representing wave propagation and accounting for mutual coupling and radiation patterns through precise modeling of the received electric field vector \cite{SangMowWin:21}. However, their complexity grows for intricate scenarios, relying heavily on numerical solutions to Maxwell's equations \cite{friedlander2019localization}. 
	
	\subsection{Algorithms for Near-Field Localization, Communications, and Sensing}  
	This subsection reviews some algorithms tailored for \ac{NF} models.  
	
	\paragraph{Localization \& Sensing Algorithms} 
	As detailed below, localization and sensing algorithms for \ac{NF} sources encounter several issues, such as high complexity due to multidimensional search and reduced accuracy due to non-Gaussianity, parameter matching, and aperture loss issues. 
	
	In particular, it is sufficient to search only over the \ac{AoA} for \ac{FF} sources. However, for \ac{NF} sources, we need to jointly search over both the angle and range because of the angle-range coupling associated with the spherical wavefront. Alternatively, the received signal can be represented in terms of two electrical angles for each source: one depends on the \ac{AoA} while the other is a function of both angle and range.  Therefore, jointly searching over these electrical angles (or over the range and angle) entails a 2D search with high computational complexity. Usually, such a search is performed using 2D  -\ac{MUSIC}, -\ac{ESPRIT}, and -\ac{CS} algorithms. Several works target splitting the 2D search into two sequential {1D} searches over the coupled parameters separately to reduce such complexity.  The computational complexity of several \ac{NF} localization algorithms is summarized in \cite[Table I]{zhang2018localization}. However, splitting the 2D-search into two 1D-search is not straightforward as first, we need to decouple the bearing angles and the distances or the electrical angles. For instance, by properly selecting some antenna elements for computing the channel covariance matrix, one of the electrical angles can be initially eliminated from the problem, allowing 1D-\ac{CS} for the \ac{AoA} \cite{rinchi2022compressive}.
	
	In \ac{NF} scenarios, sources are closer to the receiving array, leading to complex environmental interactions, including multipath propagation, diffraction, and scattering. These interactions result in non-Gaussian and non-linear signal characteristics, making traditional covariance-based methods less effective. Therefore, high-order statistics,  such as cumulants, can be useful to capture the additional structure in the signal. In this regard, cumulant-based algorithms have been proposed to sense \ac{NF} sources by decomposing the signal subspace from the noise subspace through eigenvalue decomposition of the signal cumulant rather than the typical spatial covariance matrix \cite{Challa:95}.
	
	Another issue is that for \ac{NF} scenarios with multiple sources,  matching each pair of electrical angles with their corresponding source can be challenging. Several algorithms handle this by first searching for the electrical angle, which depends only on the bearing angle. They then obtain the other angle, ensuring a proper pairing of parameters. \cite{LiagDing:10}. 
	
	Also, most \ac{NF} algorithms require the spacing between antennas to be less than a quarter wavelength for some mathematical reasons, reducing the aperture size and, consequently, the spatial resolution compared to arrays with half-wavelength spacing with the same number of elements.
	
	\paragraph{Communications Algorithms}
	The use of NF channels has great potential to enhance the capacity of the communication system \cite{dardari2020communicating}. However, these NF channels also pose specific challenges for communication algorithms due to the \ac{SNS}.  
	Specifically, the ill-conditioned nature of these channels results in slow convergence rates for low-complexity iterative algorithms designed for \ac{ZF} or \ac{LMMSE} detection performance. This is because the ill-conditioned nature of NF channels is mainly from the high level of correlation among intra-user antennas.
	User-wise (UW) \ac{SVD} approach can be used to accelerate the convergence of iterative MIMO detectors, such as Jacobi method, Gauss-Seidel (GS), gradient descent, and quasi-Newton methods in NF channels. 
	The extra computational cost of UW-SVD is worthwhile in \ac{NF} scenarios since it can accelerate the convergence of current iterative algorithms by up to ten times. However, in \ac{FF} scenarios, it can only provide similar convergence performance compared to current iterative algorithms, implying that the additional complexity introduced by UW-SVD is not justified.
	
	Beamfocusing techniques in NF demonstrate distinct characteristics compared to their FF-beamforming counterparts. While FF beamforming is primarily concerned with the angle of arrival, NF beamfocusing must also account for the unique properties of spherical wavefronts. 
	
	For instance, in the emerging network-ELAA framework, multiple distributed access points (APs) collaborate to form a functional ELAA.
	The optimal number of APs to serve the user depends on the distance between the \ac{UE} and the network-ELAA. For example, a user located at a greater distance from the ELAA would require a larger number of APs. 
	
	\section{Model Mismatch: Implications and Quantification}
	\label{sec:mismatch}
	This section discusses the implications of using incorrect field conditions, proposes tools to quantify them, and provides three case studies for multi-functional systems.
	
	\subsection{Model Mismatch in Localization and Sensing Systems}
	\subsubsection{Mismatch Challenges in Localization and Sensing Systems}
	
	The mismatch between \ac{NF} and \ac{FF} models can significantly impact distance and angle estimation in wireless localization systems. Accurate distance estimation becomes challenging when designing a system assuming a \ac{FF} model while the transmitter and the receiver are in the \ac{NF} region of each other. For example, neglecting the wavefront curvature can lead to a loss of information about the distance, i.e., the \ac{FF} array manifold discards the information about the range encapsulated in the spherical wavefront \cite{HuiChenElzanaty:22}. This can have detrimental consequences, such as positioning errors and compromised accuracy in tracking and mapping applications. 
	
	Conversely, designing localization and sensing algorithms assuming \ac{NF} conditions while operating in the \ac{FF} region may use additional degrees of freedom and features specific to the \ac{NF}. 
	This can lead to sub-optimal performance and ill-conditioned localization problems. The algorithms may struggle to estimate positions accurately, exhibit reduced accuracy in determining relative distances, and fail to account for \ac{FF} propagation characteristics. For instance, a single anchor can perform a 3D orientation estimation for sources within the \ac{NF} of the array. However, its functionality is limited to recovering only the 2D orientation angles when the source moves to the \ac{FF} \cite{EmenBuehrer:23}.
	The consequences of \ac{NF} and \ac{FF} model mismatches in localization are pervasive across various domains. These challenges underscore the importance of accurately capturing the specific field conditions and selecting appropriate models. By considering the wavefront curvature and spatial variation, system designers can mitigate the model mismatch and enhance the accuracy and reliability of localization algorithms. 
	
	\subsubsection{Metrics for Localization and Sensing Model Mismatch} \label{sec:localizaitonmetrics}
	
	In sensing and localization, different metrics quantify performance loss due to \ac{NF} and \ac{FF} model mismatches. Metrics vary in computational complexity and insights, requiring a trade-off between complexity and performance quantification. These metrics fall into three categories: \textit{i)} intuition metrics, \textit{ii)} ad-hoc metrics, and \textit{iii)} metrics for ultimate performance.
	
	\paragraph{Intuition Metrics} These metrics estimate the proximity of the widely adopted \ac{FF} model to the accurate \ac{NF} counterpart. An example is the angle between the antenna array manifold of \ac{SWM}+\ac{SNS} and that assuming a planar wavefront. Smaller angles suggest that \ac{FF} models yield comparable performance to the accurate \ac{NF} model, albeit with reduced computational complexity \cite{friedlander2019localization}. Alternatively, we can consider a \textit{statistical distance} between the distribution of measurements under the \ac{NF} model and that of the \ac{FF} model, conditioned on estimated parameters. Metrics like \ac{KL} and Wasserstein Distance fall into this category. However, these metrics generally lack precise quantification of positioning or sensing accuracy losses.
	
	\paragraph{Ad-hoc Metrics} For localization, the\ac{MSE} on the position estimate is used to evaluate the performance of algorithms under both correct and mismatched field models, while \ac{GOSPA} distance assesses mismatch losses in \ac{SLAM} and multi-target tracking. For radar sensing, the probability of detection and false alarm probability serve as performance indicators. 
	
	Considering channel sensing, it is common to assume that this covariance matrix is an identity matrix in \ac{FF}. However, this assumption becomes less valid in \ac{NF} channels, where the covariance matrix is heavily influenced by the environment.
	As a result, utilizing an identity matrix for \ac{LMMSE} channel estimation in NF scenarios can lead to suboptimal performance compared to using the exact covariance matrix.
	Therefore, a metric to quantify the mismatch can be the log-normalized difference in \ac{MSE}s for channel estimation, computed using the correct \ac{NF}  covariance matrix versus the mismatched \ac{FF} matrix. Its mathematical form is similar to that in \cite[Eq. 24]{HuiChenElzanaty:22}, where the \ac{CRLB} is replaced by the MSEs of channel estimation.
	
	\paragraph{Metrics for Ultimate Performance Bounds} In this context, relying solely on model-based statistical measures offers a more comprehensive understanding of ultimate performance loss in incorrect field assumptions, regardless of the adopted localization or sensing algorithm.
	
	In particular, mismatched error probability can be lower bounded through \ac{MLR} test, derived via \ac{ZZB} analysis \cite{xu2004bound_1}. Mismatch analysis is effectively harnessed with the \ac{ZZB}, as the \ac{MLR} test relies on both observation and statistical information assumed about it, aiding in evaluating error probability. In contrast, \ac{CRLB} integrates the observation vector, leaving only assumed statistics for expressing the bound. Thus, \ac{CRLB} proves helpful only when the estimator model aligns with data generation. However, when a model discrepancy exists, \ac{MCRLB} is necessary for accurate performance analysis.
	The \ac{MCRLB} offers a lower bound on the variance of any unbiased estimator in the presence of model misspecification, e.g., \ac{MLE} with mismatched model. Hence, it provides a lower bound on the positioning error when the measurements are obtained from the \ac{NF} model, while the estimator assumes \ac{FF}. Thus, \ac{MCRLB} offers a more insightful and accurate performance analysis of mismatched estimators that consider \ac{FF} instead of \ac{NF}. 
	
	\subsubsection{Case Studies for Mismatch Analysis in Localization and Sensing}
	We provide two use cases to show the impact of mismatched field conditions on localization and sensing.
	A brief conclusion is that the \ac{MME} becomes more significant when the users are located closer to the \ac{BS}.
	
	\paragraph{\bf Case Study 1: Implications for Single-Anchor Localization}
	\begin{figure}[t!]
		\centering
		\includegraphics[width=0.95\linewidth]{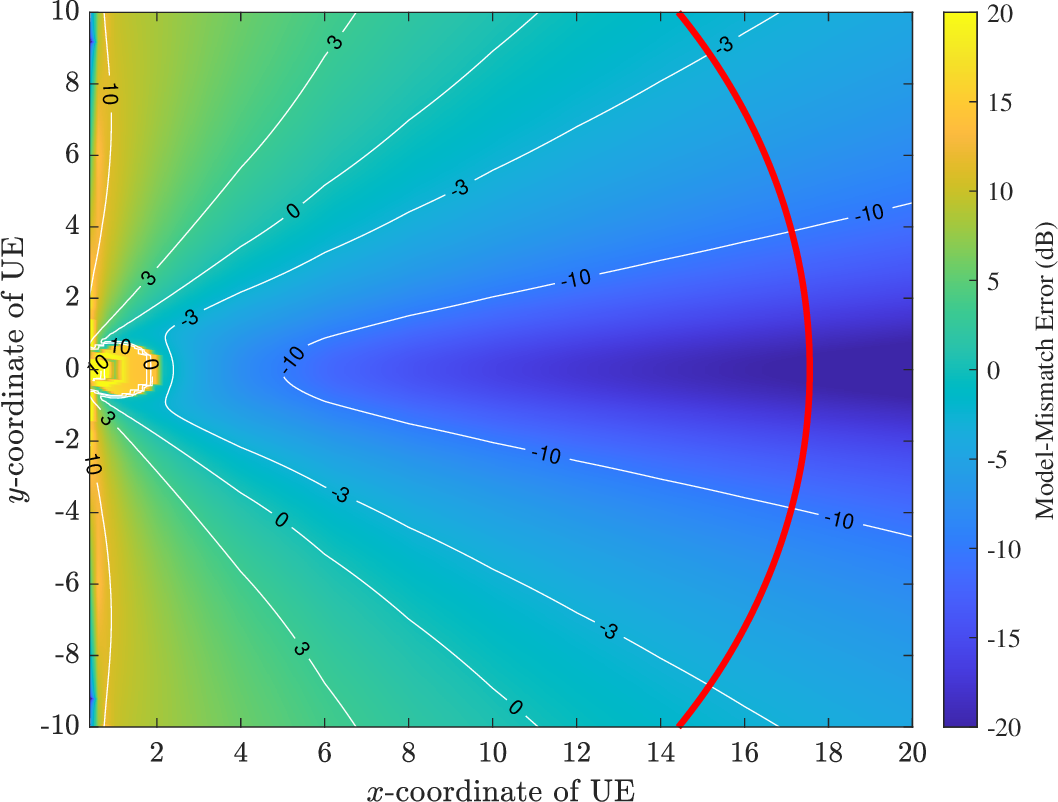}
		\caption{Visualization of the localization model mismatch errors, for different \ac{UE} positions, i.e., the normalized absolute difference between the localization error bound using the  \ac{FF} \ac{MCRLB} and the \ac{CRLB} of the true \ac{NF} model in \textrm{dB}. The red solid line indicates the Fraunhofer distance.} 
		\label{fig:MMELocalization}
	\end{figure}
	We consider a single-anchor uplink positioning scenario where a \ac{BS} estimates the position of a single-antenna \ac{UE}. 
	The \ac{BS} is located at the center and has a \ac{ULA} along the y-axis with its center at the origin with $128$ antenna elements. The channel gain depends on the distance between the \ac{UE} and the $n$-th antenna of the \ac{ULA}, and on the wavelength of the $k$-th subcarrier, capturing the \ac{SNS}.
	The \ac{UE} transmits a symbol with 10 \ac{OFDM} subcarriers at an average transmit power of 20~\textrm{dBm}, with a carrier frequency of 140 \textrm{GHz} and a bandwidth of $400$ \textrm{MHz}. The noise power spectral density and noise figure are considered as $-173.8$\,\textrm{{dBm/Hz}} and 10 \textrm{dB}, respectively. 
	
	In Fig.~\ref{fig:MMELocalization}, we show the \ac{MME} which is defined in \cite{HuiChenElzanaty:22} as the log-normalized difference between the lower bound on the localization error using \ac{MCRLB} with \ac{FF} assumption and the \ac{CRLB} of the true \ac{NF} model with \ac{SWM} and \ac{SNS}. 
	This metric quantifies the excess in the localization error bound that arises from the discrepancy in considering an estimator with a mismatched \ac{FF} model compared to that of the accurate true \ac{NF} model. Notably, the \ac{MME} is influenced not solely by distance, as indicated by the Fraunhofer distance, but also by angles. For example, the impact of model mismatch is less severe in the bore-sight of the antenna array, i.e., perpendicular to the array, even at short distances. On the contrary, the localization error due to the mismatch tends to be large for small \acp{AoA}, i.e., almost parallel to the array.  
	
	A case in point is the $3$~\textrm{dB} contour, where it becomes evident that a $3$~\textrm{dB} loss in positioning accuracy persists even beyond the Fraunhofer distance, denoted by the solid red line. Thus, relying exclusively on the Fraunhofer distance as a \ac{NF}/\ac{FF} boundary can yield misleading results.  On the contrary, opting for the more intricate \ac{SWM} when the \ac{MME} is notably small (e.g., -20 dB for the boresight of the antenna array at the Fraunhofer distance) may introduce unnecessary intricacy. This also shows that the performance of \ac{NF}-based localization algorithms behaves analogous to \ac{FF}-based algorithms when the user is in the \ac{FF}.
	
	\paragraph{\bf Case Study 2: Implication on Channel Estimation}
	\begin{figure}[t]
		\centering
		\includegraphics[width=0.96\linewidth]{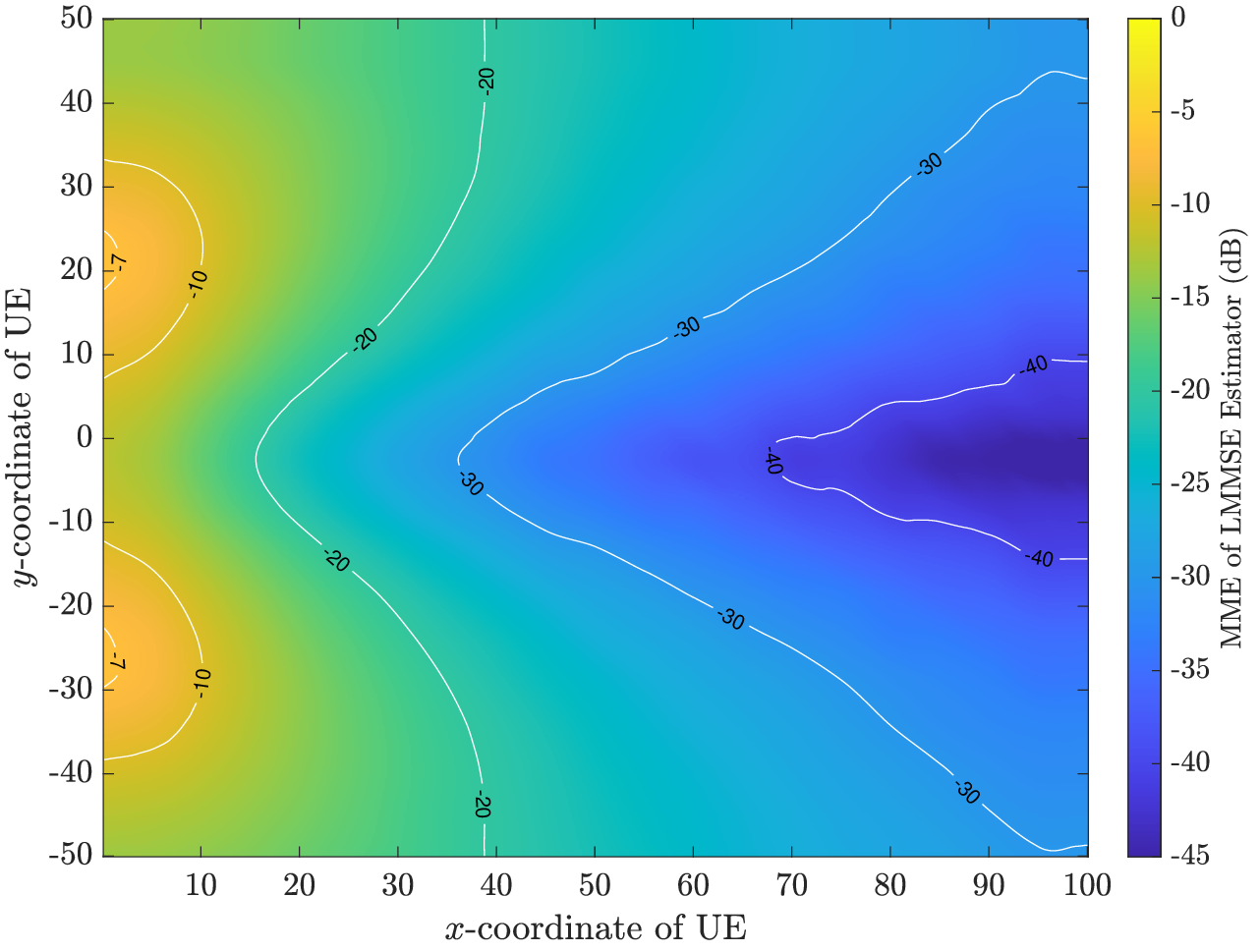}
		\caption{Visualization of channel estimation model mismatch errors, for different \ac{UE} positions, i.e., normalized relative \ac{MSE} in dB, using the  \ac{FF} identity covariance matrix and the true \ac{NF} covariance matrix inside the \ac{LMMSE}.}
		\label{fig:comm2}
	\end{figure}
	Our setup involves a BS, placed in the origin, with a large \ac{ULA} of $64$ antennas. The UE has $4\times4$ antennas with $1/3$-meter interspacing, equivalent to considering $16$ single-antenna \acp{UE}. 
	In Fig. \ref{fig:comm2}, for each location of the \ac{UE}, the color indicates the difference between \ac{MSE} of the channel estimation when \ac{NF} and \ac{FF} channel covariance matrices are considered in \ac{LMMSE}, normalized by the \ac{MSE} of the true \ac{NF}-based estimator.
	It can be observed that, for short distances between the \ac{BS} and \ac{UE}, the loss in \ac{MSE} due to incorporating mismatched \ac{FF} model can be large. However, for larger distances, e.g., more than $40\,$m, the error is negligible, especially in the foresight of the \ac{BS} antenna array. Moreover, the reduction of the \ac{MSE} loss is observed as \ac{UE} approaches the midline of the observation area. This phenomenon occurs due to the \ac{NF} channel covariance matrix approaching symmetry, resembling the characteristics of the \ac{FF} channel covariance matrix.
	\subsection{Model Mismatch in Communication Systems: Challenges, Metrics, and Use Cases}
	\subsubsection{Mismatch Challenges in Communication Systems}
	\begin{figure*}[t!]
		\begin{minipage}[b]{0.49\linewidth}
			\centering
			\centerline{\includegraphics[width=0.98\linewidth]{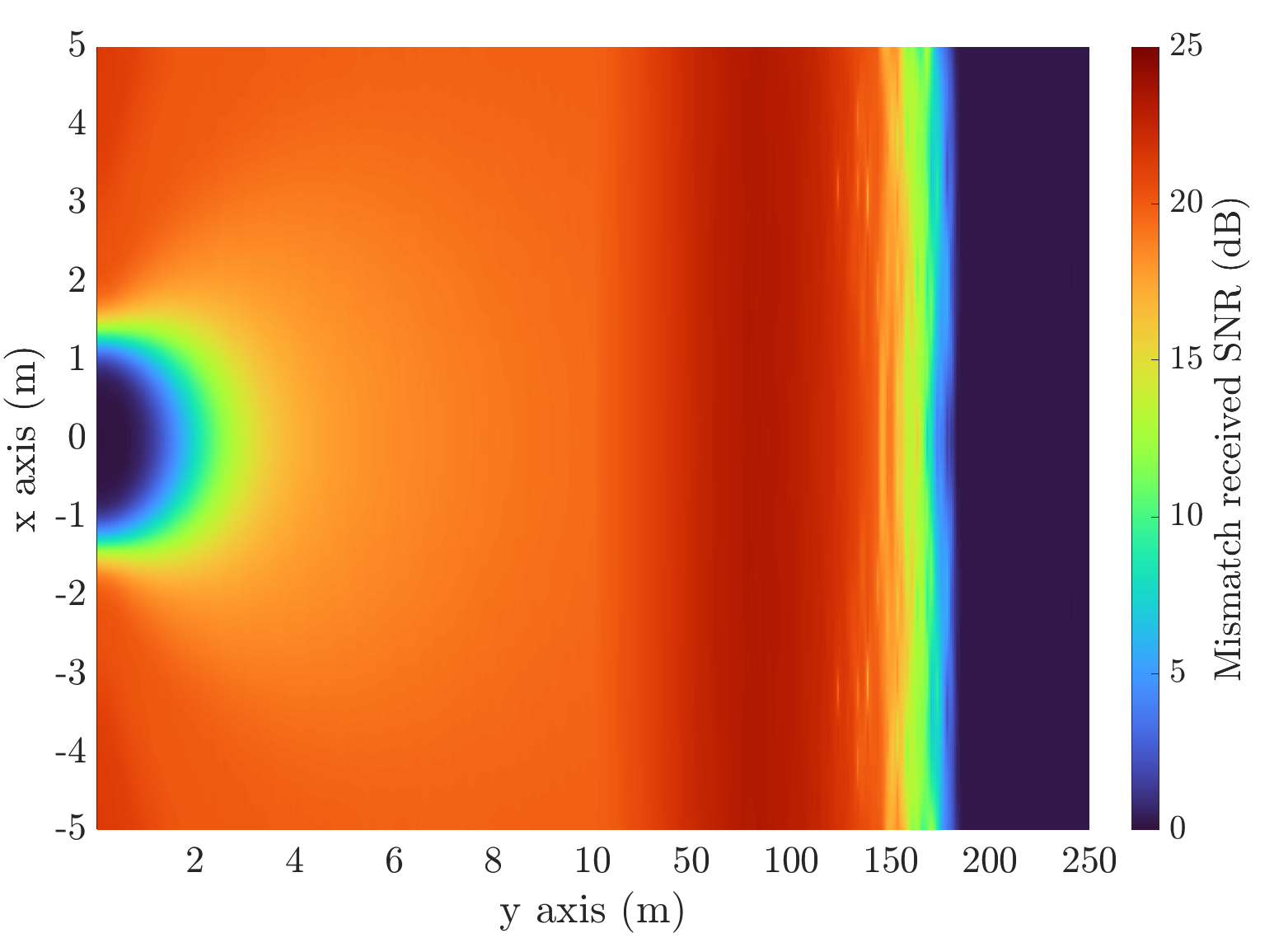}}
			\centerline{\small{(a) Aperture of ULA: 2.7m}} \medskip
		\end{minipage}
		\hfill
		\begin{minipage}[b]{0.49\linewidth}
			\centering
			\centerline{\includegraphics[width=0.98\linewidth]{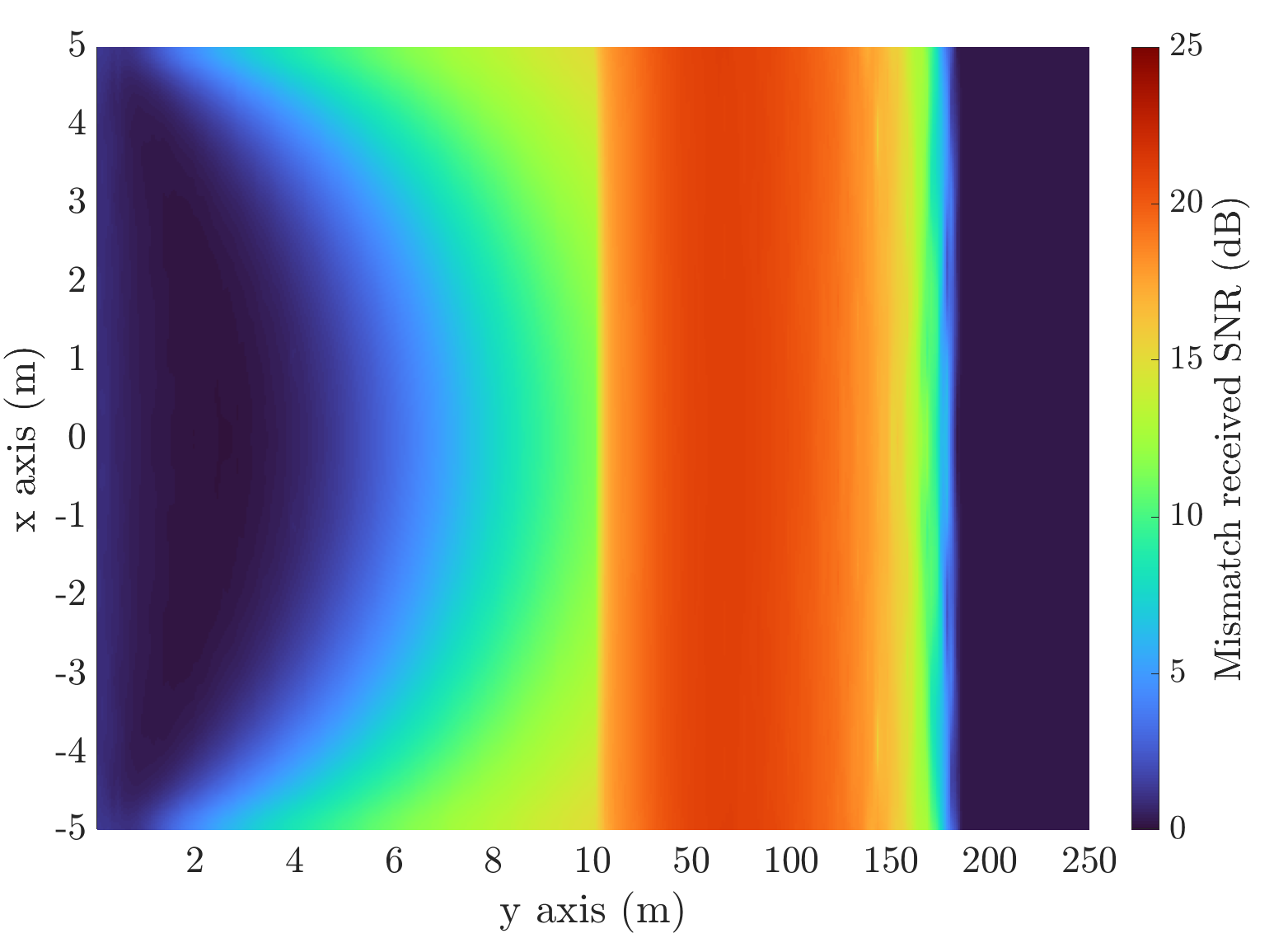}}
			\centerline{\small{(b) Aperture of ULA: 10.8m}}\medskip
		\end{minipage}
		\caption{Model mismatch of received SNR between \ac{NF} and \ac{FF} channel models to achieve a \ac{SER} of $0.001$, for different \ac{UE} positions. Note that the $y$-axis does not have a uniform scale in meters, leading to abrupt jumps in model mismatch errors at certain points.}
		\label{fig:comm1}
	\end{figure*}
	
	One of the main challenges is the use of FF models for systems primarily operating in NF conditions,  compromising beamforming effectiveness and resulting in less focused beams. Conversely, the use of NF models for FF scenarios can lead to inefficient power consumption due to the complexity of NF channels.
	
	Another issue is the sum rate, which serves as an indicator of achievable data rates. While it is generally easier to calculate the sum rate in FF systems, NF systems introduce additional complexities such as \ac{SWM} and \ac{SNS}. Overlooking these NF-specific factors can lead to inaccurate sum rate calculations, affecting the system's overall data rate.
	Inaccurate modeling can also significantly impact energy efficiency. Designing systems based on mismatched conditions can result in poor energy allocation, weaken signal strength, and undermine the system's overall performance. Furthermore, differences in energy management between NF and FF scenarios can introduce unpredictability in user experience and signal quality.
	
	Low complexity is essential for effective signal detection and precoding techniques. While iterative algorithms generally perform well in FF scenarios, their efficiency declines in NF situations due to challenging channel conditions. Therefore, new detection methods or modifications of existing algorithms tailored to NF conditions are necessary.
	
	Furthermore, accurate channel estimation presents a significant challenge, especially when it comes to determining the covariance matrices for NF MIMO channels because an imprecise covariance matrix impacts various aspects of the communication system.  
	\subsubsection{Metrics for Communication Model Mismatches}
	In this subsection, we present some useful metrics that can be used to quantify the impact of \ac{NF} and \ac{FF} model mismatch on communication systems. 
	\paragraph{Intuition Metrics} 
	The deviation of the channel covariance from the identity matrix can be used not only as a measure for sensing mismatch but also as a metric to evaluate communication performance mismatch.  Also, the \ac{SNR} and channel capacity distributions are different in \ac{NF} from those experienced in \ac{FF}. Hence, we can adopt the \ac{KL} to provide an implication about the distance between the distributions of \ac{SNR} and capacity in \ac{NF} and the corresponding statistics in \ac{FF}. Larger \ac{KL} values indicate higher model mismatch. 
	\paragraph{Ad-hoc Metrics} In FF scenarios, the performance of communication systems can be assessed using uniform reliability (i.e., experienced \ac{QoS}) metrics such as \ac{BER} and achievable rate for \acp{UE} at different locations, provided that they experience the same \ac{SNR}. However, this uniformity is not present in \ac{NF} systems due to the effects of \ac{SWM} and \ac{SNS}, where the reliability can fluctuate for \acp{UE} at different locations, even if they have the same received \ac{SNR}.
	
	Therefore, we introduce a metric to model the \ac{MME} of the received \ac{SNR}. In scenarios requiring a specific level of reliability, it is straightforward to determine the received SNR for \ac{FF} channels is straightforward. For \ac{NF} channels, the received \ac{SNR} that provides the same level of reliability can also be obtained for specific \ac{UE} locations. We then use the difference between the received \ac{SNR} values for NF and FF channels as the metric to quantify the \ac{MME}. Similarly, when the received \ac{SNR} is the same, the mismatch in reliability, such as BER or SER, can also serve as an alternative metric to assess \ac{MME}.
	
	\paragraph{Metrics for Ultimate Performance Bounds} 
	Adopting tools from information theory can help us quantify the model mismatch based on the \ac{NF} and \ac{FF} channel statistics that are independent of the specific adopted algorithms.  The difference between the ergodic capacity for \ac{NF} and \ac{FF} channels is an example of an information-theoretic metric. The mismatch can be even larger when we consider multiple access channels where the capacity region may vary significantly due to the inherent multiplexing gain for users at different focal points, even if they are in the same angular section.  
	
	An additional metric can be the loss in spatial \ac{DoF} due to the adoption of \ac{FF} instead of \ac{NF}. The \ac{DoF} represents the maximum number of parallel channels (communication modes) that can be used for communications,  and it can be quantified by the number of non-zero singular values (rank) of the channel matrix \cite{dardari2020communicating}.  The \ac{DoF} for NF channels can exceed one, even under strong \ac{LoS} conditions, while \ac{FF} experience lower \ac{DoF}, especially in non-rich scattering environments. Therefore, the difference between the \ac{DoF} in \ac{NF} and \ac{FF} can quantify the model mismatch and the loss in spatial data multiplexing gain. 
	\subsubsection{\bf Case Study 3: Implication on Received-SNR}
	We focus on an uplink \ac{MIMO} configuration featuring a \ac{BS}, placed in the origin,  with 64-element \ac{ULA}. To investigate the effect of the BS aperture, the distance between adjacent antennas is configured to be 1/2 and 2 wavelengths, resulting in ULA apertures of 2.7 and 10.8 meters, respectively. We have an \ac{UE} with four antennas, and we employ an \ac{LMMSE} detector for signal detection.  The matrix of the FF channel is assumed to follow \ac{i.i.d.} Rayleigh fading, while \ac{NF} channel has mixed \ac{LoS} and \ac{NLoS} links.
	The reason is that the aperture of a \ac{ULA} can be larger than the size of obstructing objects, which may only block a portion of the \ac{LoS} links. A statistical channel model utilizing a correlated Bernoulli distribution is proposed to describe this phenomenon of partial \ac{LoS} blockage \cite{Liu2023c}.
	In this article, we adopt this channel model to demonstrate the \ac{MME} of channel estimation, assuming that the propagation environment is an urban micro-street canyon.
	
	To quantify the mismatch, we fix the \ac{SER} to $0.001$, then compute the received-\ac{SNR} required under both \ac{NF} and \ac{FF} channels to achieve that \ac{SER}. In Fig. \ref{fig:comm1}, we report the difference between the required \ac{SNR} in dB for different locations of the \ac{UE}. In \cite{dardari2020communicating}, if the UE is located close enough to the BS antenna array, the channel column would be almost orthogonal with only LoS links. Hence, the mismatch in \ac{SNR} is small near the \ac{BS}. 
	As the distance increases, the mismatch enlarges, because the LoS links become increasingly correlated. Consequently, a higher received SNR is needed to maintain the required \ac{SER}. Ultimately, when the distance between the UE and the BS extends significantly, for example, the \ac{SNR} mismatch starts to decrease again. This reduction is due to the LoS probability approaching nearly zero when the distance exceeds.  This depends on the measurement result from the 3GPP Technical Report 38.901. Given that the MIMO channels are in an \ac{NLoS} state and the distance is sufficiently large, the mismatch between NF and FF channels becomes negligible.
	Moreover, upon comparing Fig. \ref{fig:comm1}(a) and \ref{fig:comm1}(b), it becomes evident that a larger aperture at the BS results in a more extensive area with a small \ac{SNR} mismatch. 
	
	\section{Future Directions and Emerging Technologies}
	In the following, we mention the foreseen future directions that can be undertaken and the new applications that may be affected by  \ac{NF}/\ac{FF} model mismatch. 
	
	\subsubsection{Integrated Sensing and Communication (ISAC)}
	ISAC systems combine sensing and communication on a single hardware platform to efficiently use wireless resources. This integration aims for overall gain by making the functions complement each other. However, it presents signal processing challenges, especially in maximizing degrees of freedom for both tasks and mitigating mutual interference. Dealing with \ac{NF} communication and \ac{FF} sensing, or vice versa, adds complexity, highlighting the need to understand channel characteristics and minimize \ac{NF}/\ac{FF} model mismatch interference. Addressing these challenges is essential to unlock ISAC potential in future wireless networks.
	
	\subsubsection{Intelligent Transportation Systems}
	
	An Intelligent Transportation System (ITS) refers to integrating advanced technologies and information systems into transportation infrastructure to improve the safety, efficiency, and overall effectiveness of transportation networks. A crucial component of ITS is the roadside unit (RSU), placed along roadways to facilitate communication between vehicles, infrastructure, and central control systems. 
	RSUs can collaboratively form a distributed antenna array with a large aperture, working in tandem to maximize the spatial degrees of freedom for sensing, localization, and communication. Misunderstanding of the \ac{FF}/\ac{NF} characteristics of the RSU-to-Vehicle channel could result in significant performance degradation.
	
	\subsubsection{Emerging Technologies} 
	Emerging 6G technologies, such as \acp{RIS} can manipulate electromagnetic waves, enabling field characteristic control in both \ac{NF} and \ac{FF} regions. Finally,  algorithms based on machine learning can bridge model gaps by training on diverse datasets and accommodating different propagation characteristics. 
	
	\section{Conclusion}
	This article explores the implications of \ac{NF} and \ac{FF} model mismatches in the context of 6G technology, particularly concerning communication, localization, and sensing systems. As 6G will likely operate in \ac{NF} radiating conditions, utilizing high-frequency and electrically large antenna arrays, comprehending this mismatch is paramount for system designers. The research underscores that \ac{NF} models can introduce added complexity, impacting critical facets such as localization accuracy, sensing reliability, and communication efficiency. Acknowledging these challenges enables practitioners to balance model precision and performance efficiency. Given the impact of these mismatches, it is recommended that potential design phase issues be identified.
	
	\bibliographystyle{IEEEtran}
	\bibliography{IEEEabrv,Elzanaty_bibliography.bib,LISBIB.bib,mMIMO.bib}

	\section*{Biographies}
	\begin{normalsize}
	\vspace{0.2cm}
	{\bf Ahmed Elzanaty} (S'13-M'18-SM'22) received the Ph.D. degree in Electronics, Telecommunications, and Information Technology from the University of Bologna (UNIBO), Italy, in 2018.  
	He is a lecturer at the Institute for Communication Systems (ICS), University of Surrey, UK. He has participated in several national and European projects, such as TUDOR, GRETA, and EuroCPS. 
	His research interests include wireless localization and intelligent surfaces. 
	
	{\bf Jiuyu Liu} (S'21) is currently pursuing his Ph.D. degree in Electronic Engineering with the 5GIC \& 6GIC, Institute for Communication Systems (ICS), University of Surrey, U.K. He received the M.Sc. degree in Electronic Engineering from the University of Surrey, U.K., in 2020. His main research interests include multiple-input multiple-output, extremely large aperture array, near-field channel modeling, and digital signal processing.
	
	{\bf Anna Guerra} (M'16) received the Ph.D. degree in Electronics, Telecommunications, and Information Technology from UNIBO, Italy, in 2016. She is a Researcher at CNR-IEIIT. 
	Her research interests include wireless sensor networks, radio localization, and signal processing. 
	
	
	{\bf Francesco Guidi} (M'13) received the Ph.D. degree in electrical engineering from UNIBO and the Ecole Polytechnique, Paris.
	He is now a Researcher at the IEIIT-CNR, Italy. 
	His research interests include radar networks, radio localization, and multi-antenna systems.
	
	{\bf Yi Ma} (M'04–SM'10) is a Chair Professor within the Institute for Communication Systems (ICS), University of Surrey, U.K. He has authored or co-authored 200+ peer-reviewed IEEE journals and conference papers. He holds 10 international patents in the areas of spectrum sensing and signal modulation and detection. He has served as the Tutorial Chair for EuroWireless2013, PIMRC2014, and CAMAD2015. He was the Founder of the Crowd-Net Workshop in conjunction with ICC’15, ICC’16, and ICC’17. 
	
	{\bf Rahim Tafazolli} (SM'09) is Regius Professor of Electronic Engineering, Professor of Mobile and Satellite Communications, Founder and Director of 5GIC, 6GIC and ICS at the University of Surrey. He has authored and co-authored 1,000+ publications. He was the leader of study on “grand challenges in IoT” in the UK, 2011-2012, for RCUK and the UK TSB. He holds Fellowship of Royal Academy of Engineering (FREng), Institute of Engineering and Technology (FIET) as well as that of Wireless World Research Forum. He was also awarded the 28th KIA Laureate Award 2015 for his contribution to communications technology.
\end{normalsize}
\end{document}